\documentclass[fleqn,usenatbib]{mnras}

\usepackage[T1]{fontenc}
\usepackage{ae,aecompl}

\newcommand{\mmdot}{{M\dot{M}}}

\newcommand*{\myfont}{\fontfamily{cmtt}\selectfont}
\DeclareTextFontCommand{\textmyfont}{\myfont}

\usepackage{graphicx}	
\usepackage{amsmath}	
\usepackage{amssymb}	

\usepackage{txfonts}

\title[Correlated Variability in NGC~6814]{Correlated X-ray/Ultraviolet/Optical Variability in NGC~6814}

\author[Troyer et al]{Jon~Troyer$^1$\thanks{jon.troyer@wayne.edu},
David~Starkey$^2$,
Edward~M.~Cackett$^1$,
Misty~C.~Bentz$^3$,
\newauthor
Michael~R.~Goad$^4$,
Keith~Horne$^2$,
and
James ~E.~Seals$^3$
\\$^1$Department of Physics \& Astronomy, Wayne State University, 666 W. Hancock St, Detroit, MI 48201, USA
\\$^2$SUPA, University of St Andrews, School of Physics \& Astronomy, North Haugh, St Andrews, KY16 9SS
\\$^3$Department of Physics \& Astronomy, Georgia State University, Atlanta, GA 30303, USA
\\$^4$Department of Physics \& Astronomy, College of Science and Engineering, University of Leicester, University Road, Leicester LE1 7RH}

\date{Accepted XXX. Received YYY; in original form ZZZ}

\pubyear{2015}

\begin{document}
\label{firstpage}
\pagerange{\pageref{firstpage}--\pageref{lastpage}}
\maketitle

\begin{abstract}
\noindent We present results of a 3-month combined X-ray/UV/optical monitoring campaign of the Seyfert 1 galaxy NGC 6814.  The object was monitored by {\it{Swift}} from June through August  2012 in the X-ray and UV bands and by the Liverpool Telescope from May through July 2012 in $B$ and $V$.  The light curves are variable and significantly correlated between wavebands. Using cross-correlation analysis, we compute the time lag between the X-ray and lower energy bands.  These lags are thought to be associated with the light travel time between the central X-ray emitting region and areas further out on the accretion disc.  The computed lags support a thermal reprocessing scenario in which X-ray photons heat the disc and are reprocessed into lower energy photons.  Additionally, we fit the lightcurves using CREAM, a Markov Chain Monte Carlo code for a standard disc.  The best-fitting standard disc model yields unreasonably high super-Eddington accretion rates.  Assuming more reasonable accretion rates would result in significantly under-predicted lags. If the majority of the reprocessing originates in the disc, then this implies the UV/optical emitting regions of the accretion disc are farther out than predicted by the standard thin disc model.  Accounting for contributions from broad emission lines reduces the lags in $B$ and $V$ by approximately 25\% (less than the uncertainty in the lag measurements), though additional contamination from the Balmer continuum may also contribute to the larger than expected lags.  This discrepancy between the predicted and measured interband delays is now becoming common in AGN where wavelength-dependent lags are measured.
\end{abstract}

\begin{keywords}
galaxies: active --- galaxies: individual: NGC 6814 --- galaxies: Seyfert --- accretion, accretion discs
\end{keywords}

\section{Introduction}
The current standard model of an Active Galactic Nucleus (AGN) consists of a central supermassive black hole (SMBH) actively accreting matter \citep[e.g.,][]{Rees_1984} which forms an accretion disc.  As matter is drawn toward the black hole's event horizon, gravitational potential energy is converted into kinetic and viscous internal energy.  The accretion disc then radiates thermally with the majority of the flux in the UV/optical bands \citep[e.g.,][]{troyer_12}.   X-rays from AGN are thought to be dominated by emission due to Compton up-scattering of the thermally emitted photons from the accretion disc by hot electrons in the disc's corona.  Recent measurements from X-ray reverberation and gravitational microlensing both independently imply that the X-ray emitting region is small \citep[$\lesssim$10~GM/c$^2$, e.g.,][]{demarco13, reis13,mosquera13,cackett14, blackburne15}.

In order to probe the interior structure of  AGN, a method known as reverberation mapping (RM) \citep{troyer_29} is used extensively \citep[see][for a recent review]{troyer_07}.  Reverberation mapping involves measuring the time delay associated with some variable luminosity source and the ``echo'' it produces as it interacts with matter.  Most AGN host galaxies are at distances too far for the AGN to be be spatially resolved.  In these cases, reverberation mapping provides the only direct method of probing the interior of an AGN.  In addition, reverberation mapping trades spatial resolution for time resolution.   Through reverberation mapping, the object's size scale is resolved via a time delay, i.e., the light crossing time between the source and the echo ($R \simeq c\tau $).  In principle, reverberation mapping has few limitations with respect to AGN distance as long as sufficient signal-to-noise exists, the monitoring period is long enough to detect significant variability, and the sampling is dense enough to resolve time delays between different emission components.

It has long been established that AGN spectra possess inherent variability.  A correlation between light curves of different wavelengths has been detected in many AGN \citep[e.g.,][]{Krolik_1991,Ulrich_1997,troyer_20,troyer_18,troyer_39,faus_15}.  This suggests that the emission processes associated with different wavebands are related.  If such a correlation exists for a particular object, the time lag between the X-ray and UV/optical lightcurves can be calculated in order to help understand the origin of the UV/optical variability.  There are two favored scenarios regarding the source of correlated UV/optical variability \citep[e.g][]{troyer_15, troyer_20}.  The first case is where the X-ray variability leads the UV/optical variability.  In this case it is thought that the X-ray flux heats the accretion disc and thus produces a portion of the thermal emission - the thermal reprocessing scenario.  The second case is where the UV/optical variability leads the X-ray variability.  In this case it is thought that some intrinsic thermal variability in the accretion disc exists that produces the UV/optical variability.  The UV/optical seed photons would carry their variability signature to the corona and cause the X-ray variability via Compton up-scattering. In the UV/optical leading scenario, time lags associated with the accretion disc viscous time scale would be expected.  This time scale quantifies how rapidly a perturbation in the accretion flow can propagate through the disc.  For a typical AGN supermassive black hole, the viscous timescale is of the order of months to years \citep{troyer_31}.  

Of course, it is also possible that both these scenarios are occurring simultaneously (likely on different timescales), or that other mechanisms can contribute to the lags. For instance, observations of Mrk 79 \citep{breedt09} show that on timescales of days -- weeks, the X-rays and optical bands are highly correlated, and easily explained by reprocessing, while on timescales of years there is variability in the optical not observed in X-rays, requiring an additional mechanism to produce the variations.  Similarly, in NGC 4051, while there is strong evidence for X-rays driving optical variability on short timescales (days), there is a need for another mechanism (perhaps reflected optical continuum flux from the dust torus) to account for all the optical variability observed \citep{breedt10}.  Long-term monitoring of NGC 5548 has also shown that on long ($\sim1$ yr) timescales the optical variability, while correlated with X-rays, has a higher variability amplitude.  Therefore the long-term optical variability cannot be caused by reprocessing in this case, and is more likely due to inward propagation of accretion rate changes \citep{uttley03}.  Finally, it is possible that reprocessed emission in the Broad Region (BLR) may contaminate accretion disc lags \citep[e.g.,][]{troyer_41,breedt10}.

Short timescale lags and lags that depend on wavelength are consistent with thermal reprocessing.  Here, the X-ray photons are thermally reprocessed in the accretion disc.  The simplest geometry for such a scenario is the ``lamppost'' model where the X-rays are assumed to be emitted from a centrally located point source above the plane of the accretion disc. Given the compact size of the X-ray region compared to the UV/optical emitting region, this simplification is generally agreed to be a reasonable assumption.  In the context of the lamppost model, X-ray flux is incident upon inner regions of the accretion disc before the outer regions, due to the shorter light crossing time.   See \citet{troyer_16} for a detailed description of the application of the lamppost model to continuum lags.  

NGC 6814 has been part of a previous reverberation mapping campaign \citep[the LAMP project;][]{bentz09b}.  Significant continuum variability was seen over  the approximately 70 days of monitoring, with excess variance in the B band of $F_{\rm var} = 0.18$.  An H$\beta$ lag of $\tau_{\rm cent} = 6.6\pm0.9$ days (rest frame) was measured, which, with the $f$-value from \citet{grier13} implies a black hole mass of $(1.4\pm0.3)\times10^7$~M$_\odot$ \citep{bentzDB}.  Spectroscopic monitoring from the LAMP campaign also led to measured lags in H$\alpha$, He~{\sc I}, He~{\sc II}, and H$\gamma$ \citep{bentz10}. \citet{pa14} perform dynamical modeling of the LAMP data on NGC 6814, resulting in a significantly lower black hole mass estimate of $(2.6^{+1.5}_{-1.1})\times10^6$~M$_\odot$.  Their modeling also provides an estimate of the inclination of the system of $i = 47^{+17}_{-27}$ degrees.

In this paper, we present data from a combined monitoring campaign showing short time scale ($\sim$1--3 days), wavelength-dependent time lags between the X-ray and UV/optical bands for NGC 6814 for the first time.  Observations of NGC~6814 were obtained in support of the AGN reverberation mapping campaign STARE\footnote{\url{http://www.astro.gsu.edu/STARE/}, but provided an additional opportunity to study wavelength dependent lags of the accretion disc.} In Section 2 we discuss the observations and  data reduction.  In Section 3, the lightcurve  analysis including computation of time lags between the X-ray and various wave bands, and modelling the lightcurve with a standard disc MCMC code.  Finally, in Section 4 we discuss the results of our time lag analysis and MCMC lightcurve fitting analysis and possible physical interpretations.

\section{Observations \& Data Reduction}
NGC 6814 is a Seyfert 1.5, face-on spiral galaxy with a Hubble classification of SBc and is located at $\alpha_{2000}$=+19h 42m 40.6s and $\delta_{2000}$=-10d 19m 25s and $z=0.00521$.  We use observed-frame wavelengths and flux densities in our analysis. 

\begin{figure*}
\centering
\includegraphics[width=15cm]{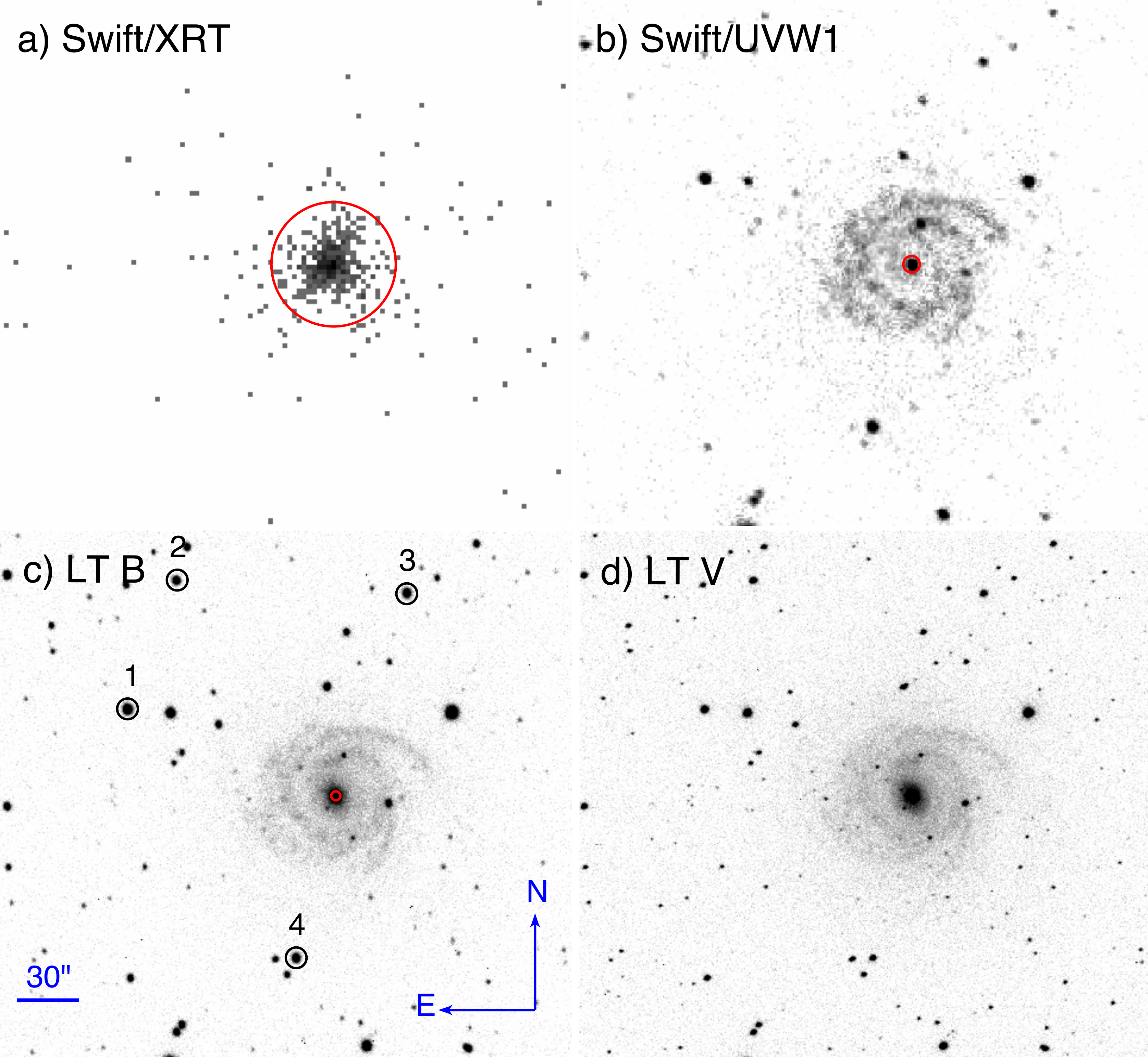}
\caption{Representative images of NGC 6814 in each waveband. (a) \textit{Swift}/XRT image when the X-ray lightcurve peaks, ObsID=00032477003, MJD 56081, with a 954 s exposure time.  For this observation, the count rate is 0.71 counts per second, corresponding to a 30\arcsec  source extraction region shown in the figure.  (b) \textit{Swift/UVW1} image, overlaid with the 4\arcsec  source extraction region used in all the observations.  This image is from ObsID=00032477024, MJD 56105, with a 663 s exposure time.  (c) \textit{LT}/$B$-band image from MJD 56092.  Black numbered circles mark the four comparison stars used in the aperture differential photometry and the red circle indicates the 2.2\arcsec extraction region used on the AGN.  The extraction region and comparison stars are common to all the \textit{LT} observations. (d) \textit{LT}/$V$-band image from MJD 56129.  }    
\label{fig:maps}
\end{figure*}

We used {\it{Swift}} \citep{troyer_22} to monitor NGC 6814 in the X-ray and UV bands.  The campaign took place over a 3-month period in 2012 resulting in 75 observations.  We also obtained optical images using the Liverpool Telescope ({\it LT}) \citep{steele04} located on the island of La Palma in the Canary Islands at the Observatorio del Roque de los Muchachos.  Representative images in each bandpass are shown in Fig. \ref{fig:maps}.  

\subsection{Swift Monitoring}

\begin{figure*}
\centering
\includegraphics[width=15cm]{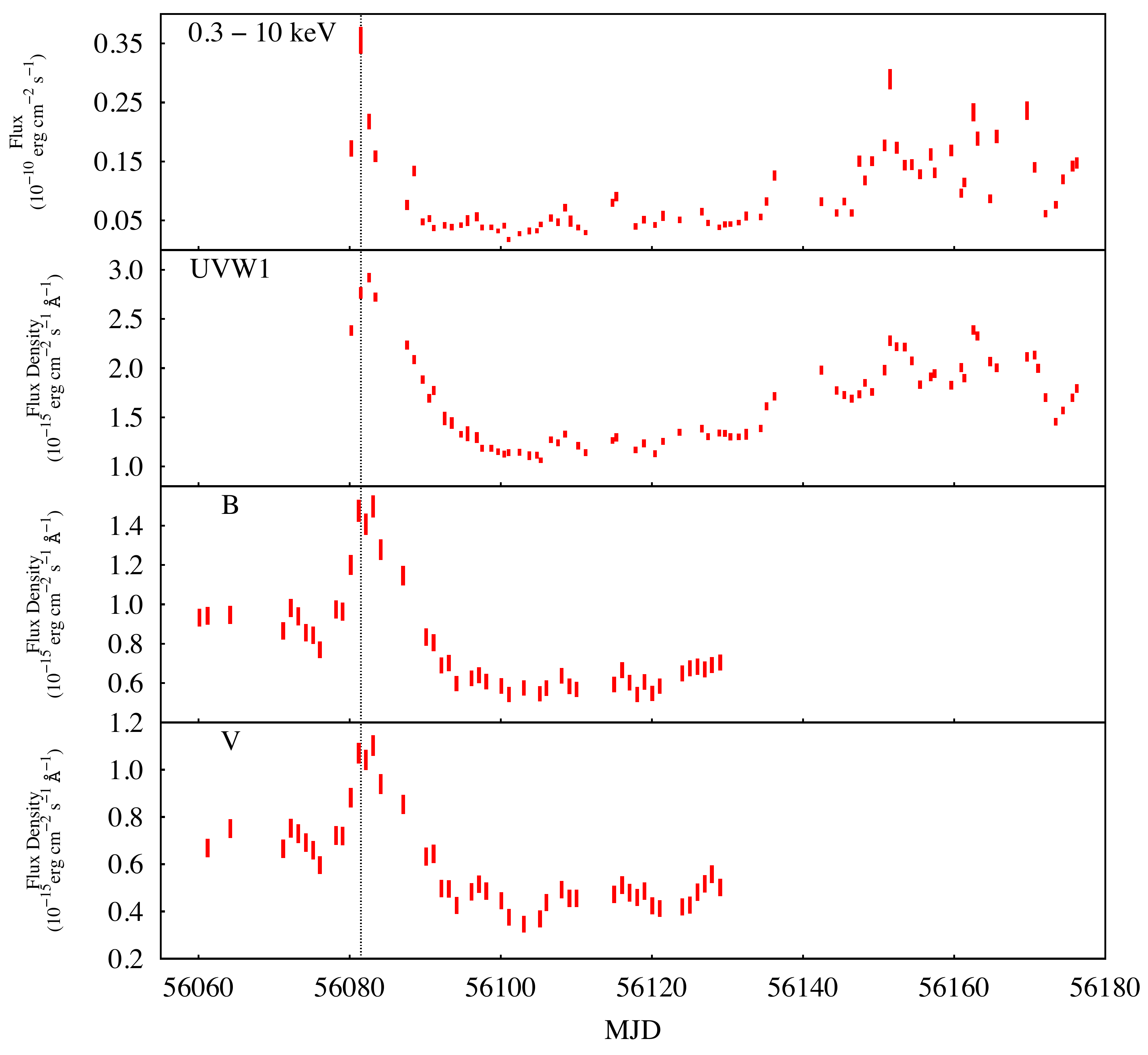}
\caption{NGC~6814 lightcurves: The top two panels show the X-ray flux  and UV flux density.  The bottom two panels are the host-galaxy subtracted $B-$band and $V-$band flux densities.  The vertical dotted line indicates the time of the X-ray flux peak.  Visual inspection shows that the strong peak in the continuum (X-ray) band is echoed in all the response bands.  Additionally, the decline from the peak in the response band lightcurves is clearly stretched with respect to the continuum, lending further support to the thermal reprocessing scenario.}
\label{fig:lcs}
\end{figure*}

NGC 6814 was monitored by {\it Swift} for a 3-month period from June 8th, 2012 until September 12th, 2012.  All dates here and throughout are in UT.  The length of the campaign and daily monitoring were selected to overlap with the concurrent STARE campaign on NGC 6814.  Nearly daily observations of ~1 ks were made with the XRT instrument \citep{troyer_24} in the $0.3 - 10$ keV energy range and UVOT instrument \citep{troyer_23}, utilizing the $UVW1$ (UV) filter, with central $\lambda =$ 2600 \AA\ and FWHM of 693 \AA.  The top two panels of Fig. \ref{fig:lcs} show the {\it Swift}  X-ray and $UVW1$ lightcurves.  Note that {\it Swift} did also obtain $V-$band images during the monitoring, however, the photometric accuracy is significantly lower than the Liverpool Telescope data, and the shape of the lightcurve was poorly constrained. We do not consider the {\it Swift} $V-$band data further.

\subsubsection{Swift X-ray Data}
We reduce the  {\it{Swift}} X-ray data using the online Build Swift XRT Products tool\footnote{\url{http://www.swift.ac.uk/user_objects}} developed by the UK Swift Science Center and described in detail in \citet{Evans_2007} and \citet{troyer_26}.  A brief overview of the data reduction follows.  We use the X-ray (0.3 keV- 10 keV) data taken in photon counting (PC) mode.  The background is calculated from an annular region around the source, and this background level is used to identify any sources detected above a 3$\sigma$ minimum.  The size of the source extraction region is selected based on the background subtracted count rate of the source, with a larger source extraction region used when the source is brighter.  The image shown in Fig. \ref{fig:maps} panel (a) is the peak of the X-ray light curve.  The count rate for this observation is 0.71 counts per second with a corresponding source extraction region of 30\arcsec.  

We convert from the XRT count rate to flux by assuming an absorbed power law model using the best-fitting parameters from \citet{walton13}, where they fit the broadband (0.5-50 keV) {\it Suzaku} X-ray spectrum of NGC~6814.  Using this model as an input, we obtain a flux conversion factor for the 0.3 keV - 10 keV band from WebPIMMS \footnote{\url{https://heasarc.gsfc.nasa.gov/cgi-bin/Tools/w3pimms/w3pimms.pl}} of 1 cps = $\text{5.0}$$\times$$\text{10}^{-11}$$\text{erg }\text{cm}^{-2}~\text{s}^{-1}$, where cps is counts per second.

\subsubsection{Swift UV Data}
We reduce the {\it{Swift}} $UVW1$ data using NASA's HEASoft\footnote{\url{http://heasarc.gsfc.nasa.gov/docs/software/heasoft/}} data analysis package.  We process the {\it{Swift}} UVOT image files with the \textmyfont{uvotbadpix} command to flag bad or damaged pixels.  Exposure map images are created, the most recent {\it{Swift}} UVOT calibration is applied and images are converted to sky coordinates using the \textmyfont{uvotexpmap} command.  Each {\it Swift} observation is often split into several shorter exposures, thus we add the various image files for each observation using the \textmyfont{uvotimsum} command.  By using the exposure map associated with each observation, all the images can be correctly oriented and summed, producing the deepest possible image.  We then use the \textmyfont{uvotdetect} command to locate any source above the detection threshold in the image. The following parameters are used: threshold=3 and chatter=5.  Searching for sources within 0.001 degrees in both RA and DEC of the known AGN location, we identify the exact location of the AGN.   We perform aperture photometry on the AGN via the \textmyfont{uvotsource} command using the \textmyfont{uvotdetect} source position. We take the source extraction region as a circle centered on the AGN, with a radius of 4\arcsec.  This region is shown in Fig. \ref{fig:maps}, panel (b). We estimate the background rate from an annular region around the AGN with an inner radius of 6\arcsec and an outer radius of 9\arcsec.  We use the following parameters: sigma=3, chatter=1, apertcorr=CURVEOFGROWTH.  We perform the same procedure for all {\it Swift} observations in order to create a lightcurve.  Flux conversion for $UVW1$ \citep{troyer_23} is 1 cps = $\text{4.3}$$\times$$\text{10}^{-16}$$\text{erg }\text{cm}^{-2}\text{s}^{-1}\text{\AA}^{-1}$.

\subsection{Liverpool Telescope Observations}
NGC 6814 was monitored by the Liverpool Telescope from May 12th, 2012 to July 20th 2012.  The observations used the RATCam instrument, operated with $2\times2$ pixel binning, which leads to a pixel scale of  0.277\arcsec\ per binned pixel, and $1024\times1024$ pixel images.
Observations were taken in pairs of exposures for each of the two filters used, $Bessel$ $B$ and $Bessel$ $V$, on a nearly daily basis.  A total of 92 pairs were taken over the roughly 2-month campaign.  Apart from the first 5 exposures which were single exposures, 45 seconds in length, the remaining pairs of exposures were 60 seconds per exposure (120s total). 

\subsubsection{Aperture Photometry}
We perform aperture photometry on the AGN and comparison stars using a circular aperture with an 2.2\arcsec (8 pixel) radius, shown in Fig. \ref{fig:maps}, panel (c).  The aperture size is based on the seeing values.  The mean seeing (FWHM) during the observations is 1.45\arcsec (5.2 pixels), with 90\% of the observations having seeing FWHM less than the aperture. The sky background is determined from the mode of values within an annulus with inner and outer radii of 4.2\arcsec (15 pixels) and 5.5\arcsec (20 pixels) respectively.    The data are typically obtained as a pair of exposures taken sequentially.  Thus, to maximize the signal-to-noise ratio we average the count rates between pairs of exposures.  

We choose four comparison stars of comparable brightness to NGC~6814, shown in Fig. \ref{fig:maps}, panel (c).  We perform differential photometry by calculating the average scale factor for each observation for the four comparison stars, assuming that they remain constant over time. We then apply this scale factor to NGC 6814 to recover the AGN lightcurve.  We get standard deviations of 0.3\%, 0.7\%, 0.7\% and 0.5\% for the four comparison star lightcurves in the $B$-band, and 0.4\%, 0.3\%, 0.6\% and 0.8\% in the $V$-band.  We find that the AGN lightcurve has a standard deviation of 11\% in the $B$-band and 6\% in the $V$-band, indicating significant variability.

A lower limit to the fractional uncertainty on the AGN count rates is 0.8\% from the highest standard deviation of the comparison stars.   As another estimate of uncertainties in the AGN count rates, we look at the difference in rate between observations that are 1 day apart.  We find the median difference to be 1.9\% for the $B$-band, and 1.1\% for the $V$-band, and we adopt these as the fractional uncertainties.  This gives an upper limit on the uncertainty, since there will likely be some real variability on this timescale.

We convert from relative rates to flux by obtaining the $B$- and $V$-band magnitude of the brightest comparison star, Star 1 shown in Fig. \ref{fig:maps}, panel (c), by using {\it HST} photometry from data in \citet{Bentz_2013} to calibrate the $V-$band photometry of our image.  This yields  a Star 1 magnitude of 14.4 in the $V-$band, which differs slightly from the SIMBAD value of 14.2.  As a check, we verified that this method recovers the published magnitudes of reference stars in \citet{doroshenko05}.  For the $B-$band, where {\it HST} photometric calibration was unavailable, we used \citet{doroshenko05} stars to calibrate our $B-$band image, yielding a Star 1 magnitude of 15.1, which is in agreement with the the SIMBAD value.  We used the zero points for Vega fluxes from \citet{troyer_36}.

\subsubsection{Host Galaxy Flux}
In order to accurately quantify the AGN variability and flux obtained from the aperture photometry,  we carry out subtraction of host galaxy light in the visual bands using methods detailed in \citet{Bentz_2006,Bentz_2009}.  Using an {\it HST} image of NGC~6814 (WFC3, F547M filter), with the AGN PSF and the sky subtracted \citep{Bentz_2013}, we duplicate the circular aperture and its background annulus (which would include some host-galaxy light) and measure the amount of host flux.  In the F547M filter, the host galaxy flux is $\text{2.7}$$\times$$\text{10}^{-15}$$\text{erg}\;\text{s}^{-1}\text{cm}^{-2}\text{\AA}^{-1}$. Assuming a typical bulge template \citep{Kinney_1996}, and using \textmyfont{Synphot}\footnote{\url{http://www.stsci.edu/institute/software_hardware/stsdas/synphot}} to carry out synthetic photometry, we estimate a $B-$band host-galaxy contribution of $1.5\times10^{-15}$ erg s$^{-1}$ cm$^{-2}$ \AA$^{-1}$ and a $V-$band host-galaxy contribution of $2.6\times10^{-15}$ erg s$^{-1}$ cm$^{-2}$ \AA$^{-1}$.
 
\subsubsection{Difference Imaging Photometry}
For comparison with the aperture photometry, we also derive the $B-$ and $V-$band light curves by registering each set of images to a common alignment using {\tt Sexterp} \citep{siverd12} and then applying the image subtraction software package {\tt ISIS} \citep{alard98,alard00}.  {\tt ISIS} builds a reference frame from the images that have been defined by the user to have the best seeing and lowest background levels.  This reference frame is then convolved with a spatially-variable kernel to match the point spread function of each individual image in the set.  Subtraction of the frame from the convolved reference image results in a residual image where the only sources are regions of variable flux.  The lightcurve is then derived from aperture photometry that is carried out on these residual images.  All contributions from constant-flux components, such as an AGN host galaxy, are thus naturally removed. 

To convert the image-subtraction lightcurves from units of residual counts to calibrated fluxes, it is necessary to know the magnitude of the source in the reference frame.  We determine this by modelling the $B-$ and $V-$band reference frames with {\tt Galfit} \citep{peng02,peng10}.  We first build a model point spread function for each frame by fitting three Gaussians to a non-saturated and well-isolated field star.  We model the entire frame in each band with the host-galaxy geometric parameters held fixed to the values determined from a high-resolution {\it HST} image by \citet{Bentz_2013}, but scaled to the appropriate plate scale.  This method results in a clean subtraction of the main host-galaxy features and allows us to accurately separate the host-galaxy flux from the AGN flux in the reference images.  By including a field star with known magnitudes in the modelling, we are able to simultaneously solve for the photometric solution in each bandpass.  Once we determine the reference AGN flux in each band, we then convert the lightcurves from residual counts to calibrated fluxes.

We found that  the fluxes derived from difference imaging are in excellent agreement with the host galaxy subtracted aperture photometry results.  We therefore use the aperture photometry results throughout the rest of the analysis.

\section{Data Analysis}
\begin{figure}
\centering
\includegraphics[width=8cm]{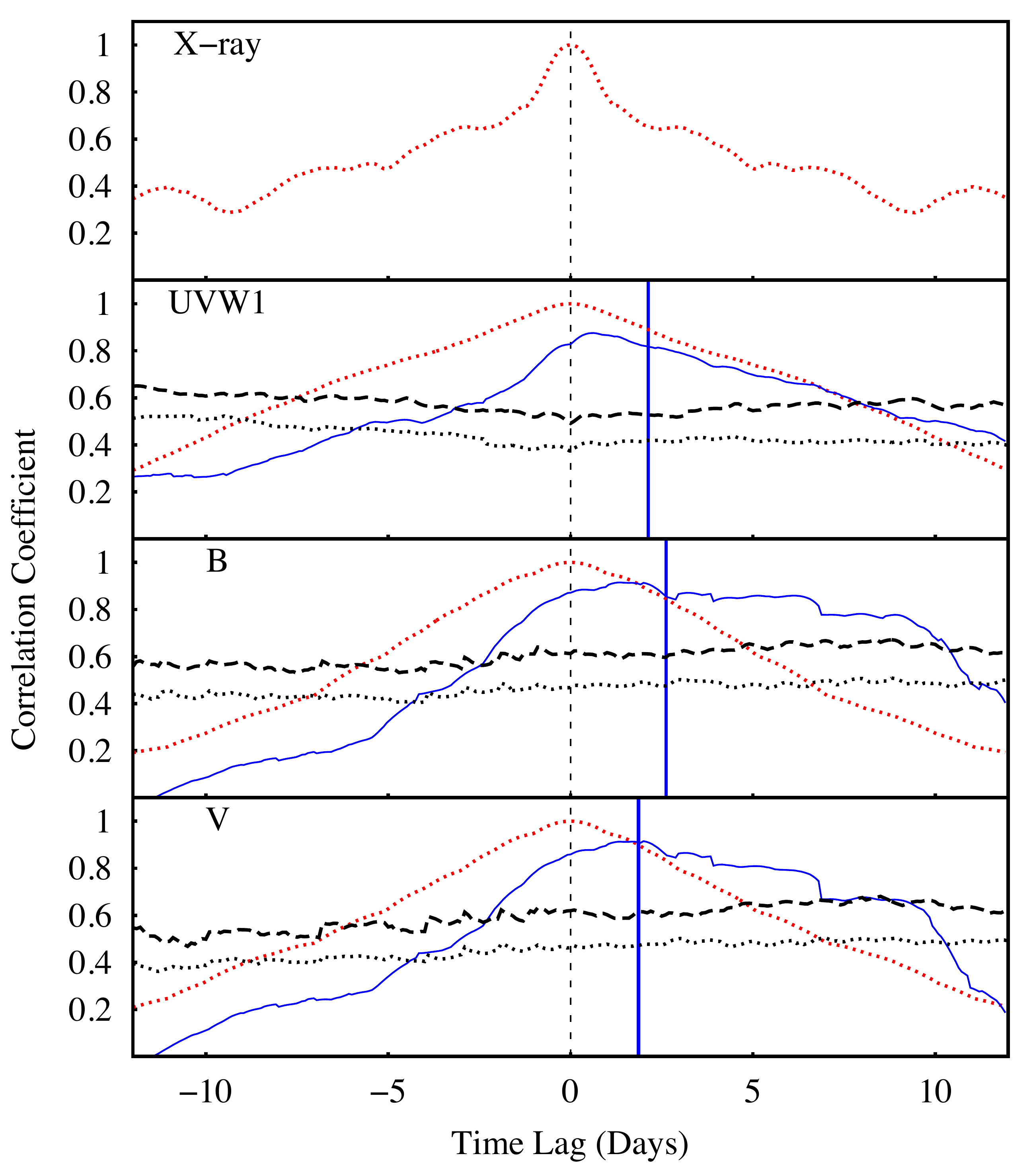}
\caption{The auto-correlation function of the X-ray lightcurve is shown in the top panel. In the lower panels, the cross-correlation functions of each band with respect to the X-ray lightcurve and auto-correlation functions are shown in increasing wavelength order.  Blue solid lines show the CCFs while the red dotted lines show the ACF of each band.  The centroid of the lags are shown by the vertical solid blue lines. The centroid values are listed in Table \ref{tab:LagTable}.  The 95\% and 99\% confidence limits in the CCF values are shown as black dotted and dashed lines respectively.}
\label{fig:xuvccf}
\end{figure}

The time lags between wavebands are quantified by using the cross-correlation function (CCF) as described in \citet{troyer_32}.  We calculate three CCF($\tau$), one for each of the response bands: UV, $B$-band, and $V$-band.  For each CCF($\tau$), we take the X-ray lightcurve to be the driving lightcurve and set the UV, $B$-band, and $V$-band lightcurves as the responding lightcurve.  For each CCF calculation, we interpolate the two lightcurves in order to obtain regular sampling.  In this fashion, the CCF values are computed twice.  The first by interpolating the continuum lightcurve so as to pair up all the continuum data points with the data points of the responding lightcurve.  The second CCF value is computed in the same way, except the responding lightcurve is interpolated as to pair up with the continuum lightcurve.  The two CCF values are then averaged at each time, yielding CCF($\tau$).  To avoid needing to extrapolate the lightcurve data, the CCF sum is restricted to the intersection of the time intervals covered by the driving lightcurve and the shifted echo (response) lightcurve. For our data, the centroid is calculated using points above 80\% of the maximum value.  The CCF($\tau$) plots are shown in Fig. \ref{fig:xuvccf}.  Additionally, we computed the auto-correlation function (ACF) for each lightcurve.  The full width, half maximum values for the ACFs are 4.7 days for X-ray, 9.1 days for UV, 6.4 days for $B$-band, and 6.8 days for $V$-band.  The ACFs are also shown in Fig. \ref{fig:xuvccf}. 

To determine confidence limits on the significance of the CCF values we follow the method of \citet{breedt09}.  We simulate an X-ray lightcurve 10 times the length of the observing campaign using the algorithm of \citet{TK_1995}.  We assume a power-density spectrum with slope of $-1$ breaking to a slope of $-2$ at frequencies above the characteristic break frequency.  We determine the break frequency by assuming it scales with mass and Eddington fraction, following \citet{McHardy_2006}, and assuming the black hole mass from \citet{pa14}, and Eddington fraction of 0.01.  We sample the simulated X-ray lightcurve at the same time intervals as the real \textit{Swift} lightcurve, and add random Gaussian noise based on the fractional uncertainties of the real data.  We then calculate the CCF between the simulated X-ray lightcurve and the $UVW1$, $B-$ and $V-$band lightcurves in turn.  We perform this 1000 times and use the distribution of CCF values at each lag to determine the 95\% and 99\% confidence levels, shown as dotted and dashed lines in Fig. 3.  The observed CCFs all peak above the 99\% level, showing that the correlations are highly significant. 

\begin{figure}
\centering
\includegraphics[width=8cm]{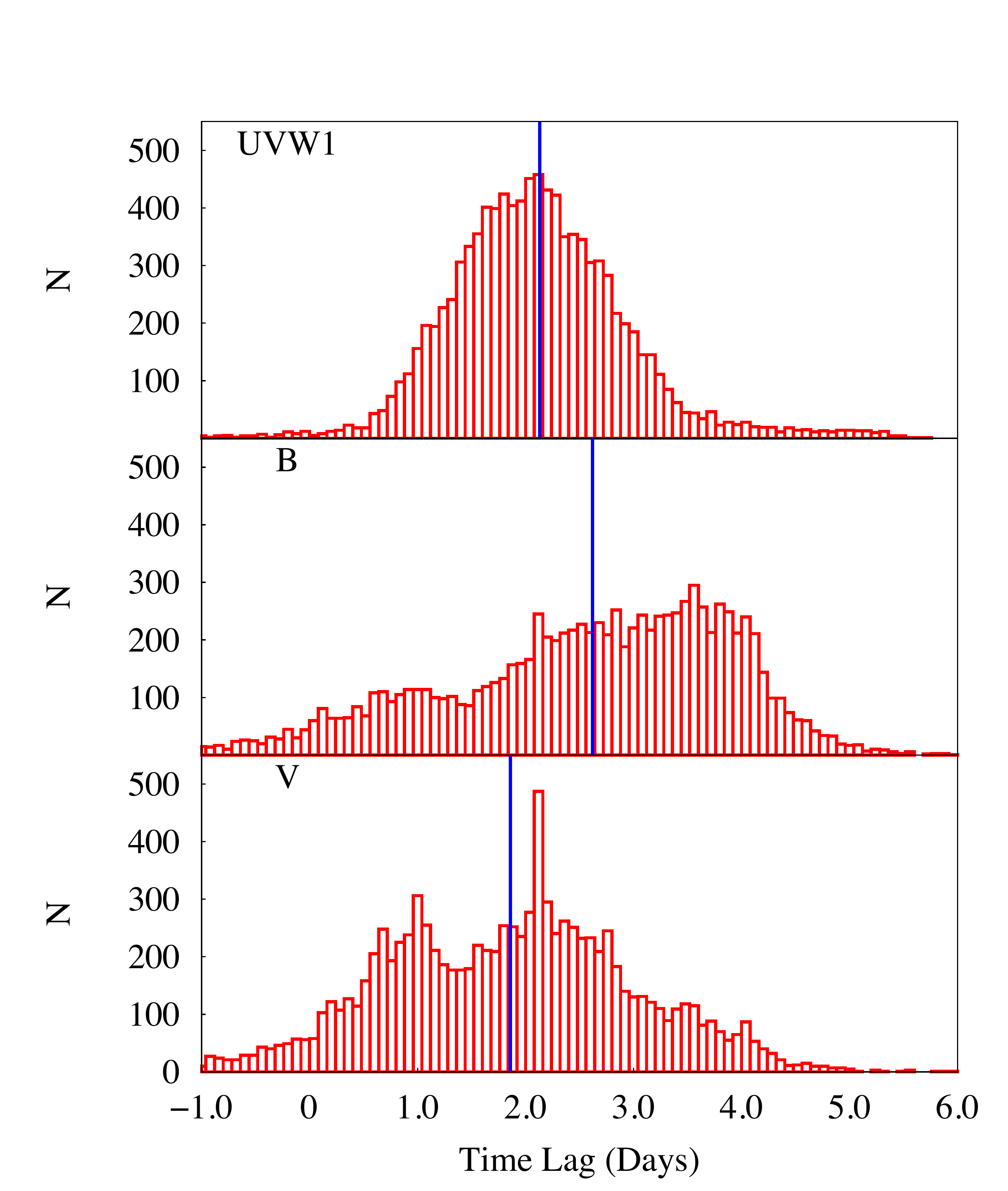}
\caption{The histogram of lag centroids for each band.  The lightcurve data were randomly resampled and the CCFs computed for 10000 realisations.  This Monte Carlo method allows us to estimate the uncertainty in the lag calculation \citep{troyer_33}.}
\label{fig:laghist}
\end{figure}

In order to quantify the uncertainty of our time lags we use Monte Carlo and Bootstrap techniques to resample and randomize the data points on each lightcurve. See \citet{troyer_33} for a discussion of CCF uncertainties. For each point, we add random Gaussian noise based on the uncertainty associated with the count rate measurement for that point.  We then randomly resample the data (Bootstrap) with the temporal ordering intact, but allow for the possibility of sampling a particular data point more than once while keeping the same number of elements of the data set, i.e., some points were excluded.  We repeat the random resampling and compute the CCF($\tau$) for each.  This is done for 10000 realisations, allowing us to build a histogram of the centroid of the CCFs, which we show in Fig. \ref{fig:laghist}.  We take the mean of the distribution of centroids as the lag value ($\tau$).  The uncertainty in the lag is taken at the 1$\sigma$ value of the distribution.  The values of time lags associated with $UVW1$, $B$-, and $V$-band lightcurves are shown in Table \ref{tab:LagTable}.  These data show the time delay between the X-ray and longer wavelength lightcurves. 
  
\subsection{Cross-correlation Lag Results}
In order to conduct accurate analysis of the time lags of the response bands, several factors are needed.  First, the dense monitoring campaign we undertook gave us the well-sampled data quality required to limit the uncertainties associated with the time delay.  Second, the intrinsic variability of the source lightcurve and the corresponding correlated response must also exist.  This  allows for the computation of the CCF($\tau$) and the time lag. Indeed, the greater the variability, the more accurately we can compute the CCF($\tau$).  One measure of the intrinsic variability is the fractional root-mean-square variability amplitude or F$_{var}$ which is described in  \citet{troyer_34}.  This statistic is computed by subtracting the variance in the individual count rate measurement errors from the variance of the count rates themselves.  This difference is called the excess variance.  
\begin{table}
\caption{Time Lags}
\centering
\begin{tabular}{c|c}
\hline
Response Band & Time Lag (days) \\  \hline
UV (2600 {\AA}) & $2.1\pm0.7$ \\ 
$B$ (4400 {\AA}) & 2.6$^ {+1.3}_{-1.5}$\\ 
$V$ (5500 {\AA}) & 1.9$^{+1.1}_{-1.2}$\\ \hline
\end{tabular}
\label{tab:LagTable}	
\end {table}

F$_{var}$ is the normalized expression of the excess variance.  The errors in F$_{var}$ are computed assuming errors only due to Poisson noise.  See Appendix B of \citet{troyer_34} for a discussion.  Values of F$_{var}$ are listed in Table \ref{tab:FvarTable} and provide a metric for measuring variability.  In the $B-$ and $V-$bands, we calculate it using the host-galaxy subtracted fluxes.

\begin{table}
\caption{Fractional Variability}
\centering
\begin{tabular}{c|c}
\hline
Band  & F$_{var}$ \\  \hline
X-ray (8.3 {\AA}) & 0.70$\pm{0.01}$ \\ 
UV (2600 {\AA}) & 0.267$\pm{0.002}$ \\ 
$B$ (4400 {\AA}) & 0.323$\pm {0.008}$\\ 
$V$ (5500 {\AA}) & 0.320$\pm{0.010}$\\ \hline
\end{tabular}
\label{tab:FvarTable}	
\end {table}

Visual inspection of the lightcurves shown in Fig. \ref{fig:lcs} indicates good correlation of all bands, as is also apparent from the peak values of the CCFs.  An initial large peak in the X-ray LC that is echoed in all the responding bands can be seen.  Moreover, each of the longer wavelength responding bands shows a broader peak as expected if the continuum is thermally reprocessed -- reprocessing on the near-side of the disc will be seen before reprocessing on the far-side of the disc -- blurring out the sharp peak seen in X-rays. 

Inspection of Fig. \ref{fig:xuvccf} reveals a moderately flat CCF($\tau$) for the $B$-band and $V$-band.  The large uncertainties in these lags arise from lack of significant overlap of the {\it{Swift}} and {\it{LT}} data as well as a period of low variability in the flux across all bands shorty after the large rise seen at the beginning of the monitoring period.  As a test, we also carried out the lag analysis with only the overlapping portion of the LCs.  We found the differences in lag distributions to be negligible. 
  
Our data support thermal reprocessing with an X-ray to UV lag of  $\text{2.1}^{+0.7}_ {-0.7}$ days, an X-ray to $B$-band lag of $\text{2.6}^ {+1.3}_{-1.5}$ days, and an X-ray to $V$-band lag of $\text{1.9}^{+1.1}_{-1.2}$ days. Thermal reprocessing of the X-ray continuum would result in wavelength-dependent time lags: $\tau$ $\propto$ $\lambda^{\frac{4}{3}}$\citep{troyer_25}.  For a standard thin disc \citep{S_S}, the temperature profile is given by \citep[e.g.,][]{troyer_25,troyer_35,troyer_16}:
\begin{equation}
T(R) =\left[\frac{3GM\dot{M}}{8\pi R^3\sigma} + \frac{L_{x} H_{x}(1-A)}{4\pi R_{x} ^3 \sigma}\right]^\frac{1}{4}\, , 
\label{eq:T(R)}
\end{equation}
where $G$ is Newton's universal gravitational constant, $M$ is the mass of the black hole, $\dot{M}$ is the mass accretion rate, $\sigma$ is the Stefan-Boltzmann constant, $L_{x}$ is the luminosity of the continuum irradiating source, $A$ is the disc albedo, $R_{x}$ is the distance from the irradiating source to the disc element at distance $R$ from the black hole, and $H_{x}$ is the height of the irradiating continuum source above the disc.
The first term in the temperature profile equation is the contribution of the viscous heating of the disc and is valid for $R\gg R_{*}$, where $R_{*}$ is the innermost stable orbit of the blackhole.  The second term is the contribution associated with radiative heating of the disc.  In the same regime: $R\gg R_{*}$ and when $R\gg H_{x}$, the second term $\propto R^{-\frac{3}{4}}$.  Overall, this suggests that $T(R) \propto R^{-\frac{3}{4}}$.  If we assume a Wien's Displacement Law relationship ($\lambda \propto T^{-1}$) for each disc element a distance $R$ from the central black hole and considering the previous relationship $T(R) \propto R^{-\frac{3}{4}}$, we obtain the relationship $R^{-\frac{3}{4}}\propto\lambda^{-1}$.   Assuming $R \simeq c\tau$ we obtain $\tau$ $\propto$ $\lambda^{\frac{4}{3}}$.

\begin{figure}
\centering
\includegraphics[width=8cm]{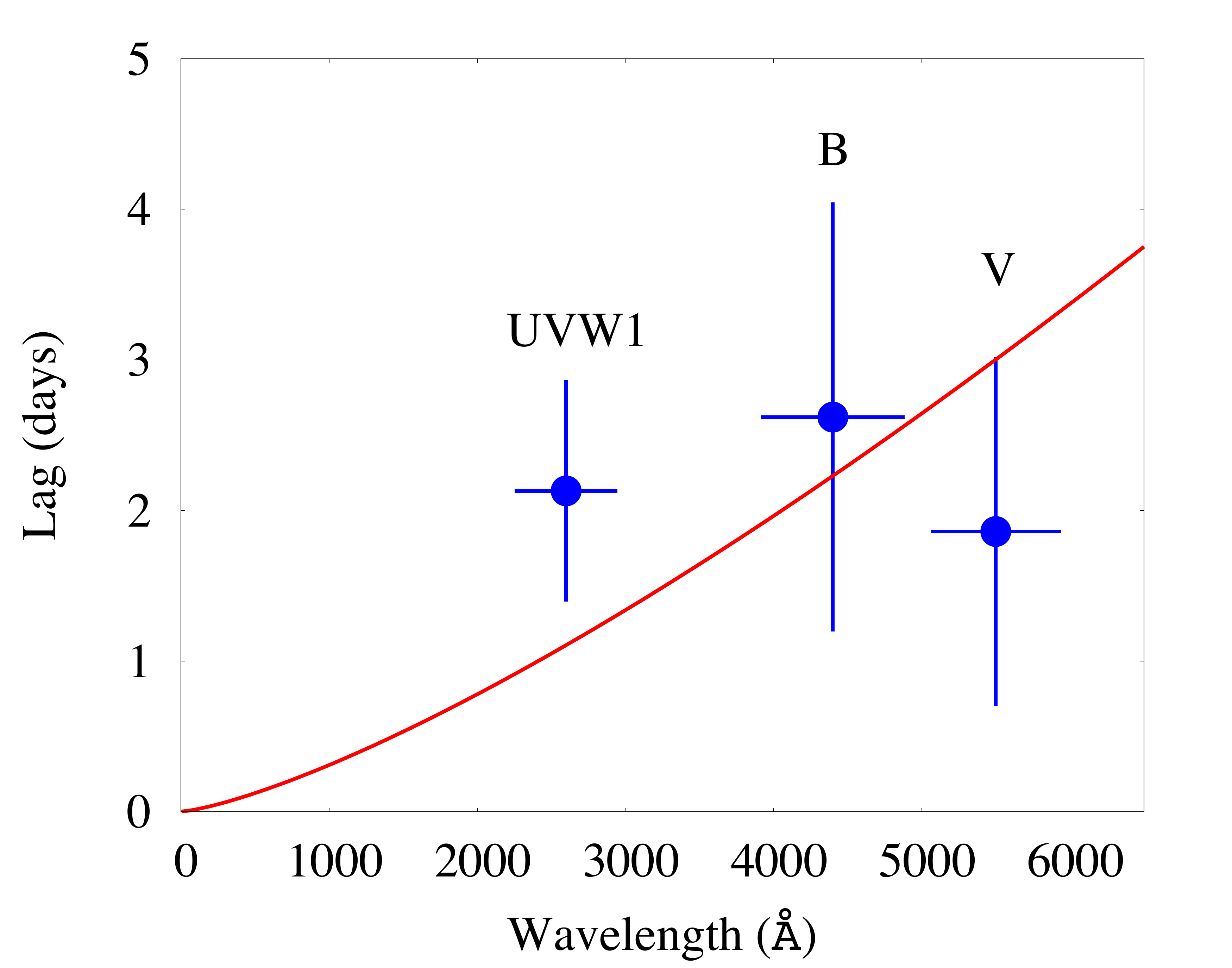}
\caption{Time lags  for the $UVW1$, $B$- and $V$-band calculated with respect to the X-ray band as a function of wavelength.  The red line is the best-fitting $\tau \propto \lambda^{4/3}$ relation, showing the data is broadly consistent with thermal reprocessing.  The x-error bars indicate the filter bandpass HWHM, so together they show the FWHM.}
\label{fig:wavelag}
\end{figure}

In Fig. \ref{fig:wavelag}, we plot time lag vs. wavelength for the $UVW1$, $B$-band, and $V$-band wavebands relative to the X-ray band.  Additionally, we plot the function:
\begin{equation}
\tau = \tau_{0}\left[\left(\frac{\lambda}{\lambda_{0}}\right)^{\alpha}-1\right]\, , 
\label{eq:lag}
\end{equation} 
where: $\lambda_{0}$ is the wavelength of the driving X-ray band (here we use a value of $\lambda_{0} = 8.3\;\AA$), $\tau_{0}$ is the continuum reference time, determined by fitting the data, and $\alpha$ is the characteristic exponent.  Fixing, $\alpha=4/3$, the relation fits the data well, but given the large uncertainties in the $B-$band and $V-$band lags, we cannot better constrain the exact wavelength dependence of the lags. 

\subsection{Monte Carlo Accretion Disc Lag Distribution Analysis}
We now perform an additional analysis of the lightcurve lags. We use the accretion disc modelling code CREAM \citep{starkey15} to fit a lamp-post model \citep[e.g][]{troyer_25,troyer_16,ch13} to the continuum emission; interpreting this as variable black body emission from a standard thin disc. 

\begin{figure*}
\centering
\includegraphics[width=13cm]{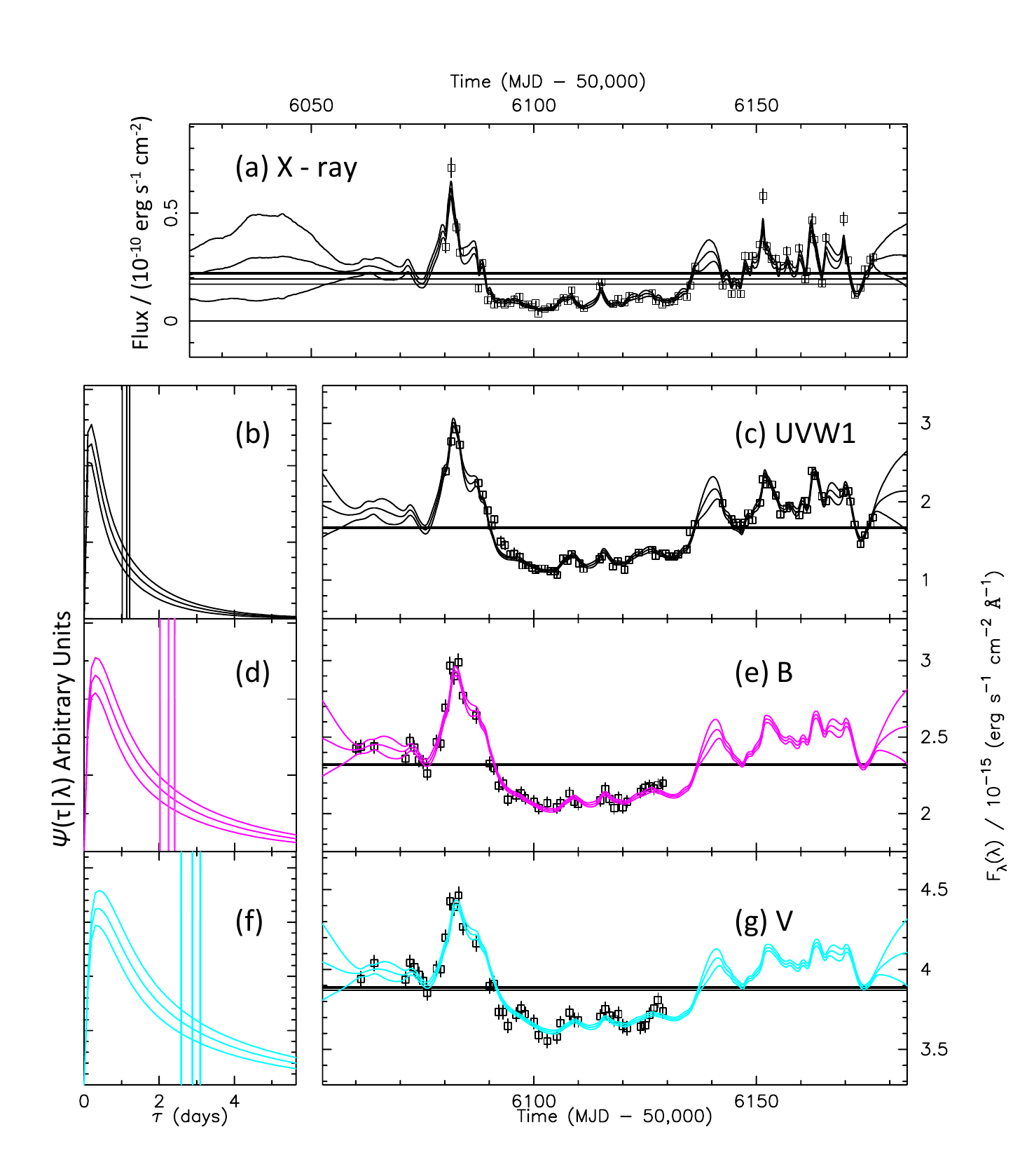}
\caption{CREAM fit for i =50$^\circ$ to the X-ray (a), {\it UVW1} and {\it LT} light curves (lower right, panels c, e and g). CREAM assumes the X-ray light curve drives the variability at the longer wavelengths and attempts to infer the disc response function (lower left, panels b, d and f).  The vertical lines indicate the mean lag and 1-$\sigma$ uncertainty envelope.}
\label{fig_50deg}
\end{figure*}

CREAM uses Markov Chain Monte Carlo (MCMC) methods to fit a simple irradiated disc model to the observed lightcurves. The driving (X-ray) lightcurve is modelled as a Fourier time series in $\log_{10}F_\lambda$, with a random walk prior on the Fourier amplitudes. Each echo (UV and optical) lightcurve is modelled as a constant flux plus variations obtained by convolving the driving lightcurve with the time delay distribution appropriate for a flat steady-state blackbody accretion disc irradiated by a variable point source just above the disc centre.

The MCMC fit samples the joint posterior probability distribution of the model parameters. The parameters of primary interest are $M\dot{M}$, which controls the $T(r)$ profile of the disc, and the disc inclination $i$. The $\mmdot$ estimate maps directly onto a mean delay with a theoretical scaling of $\langle \tau \rangle \propto (\mmdot)^{1/3} \lambda^{4/3}$ \citep{troyer_38}, independent of $i$, and the shape of the delay distribution depends on $i$. The model has hundreds of nuisance parameters, including the Fourier amplitudes that define the X-ray lightcurve, and a mean and RMS amplitude and an error bar scale factor for each echo lightcurve. For further details see \citet{starkey15}.

\begin{figure}
\centering
\includegraphics[width=8cm]{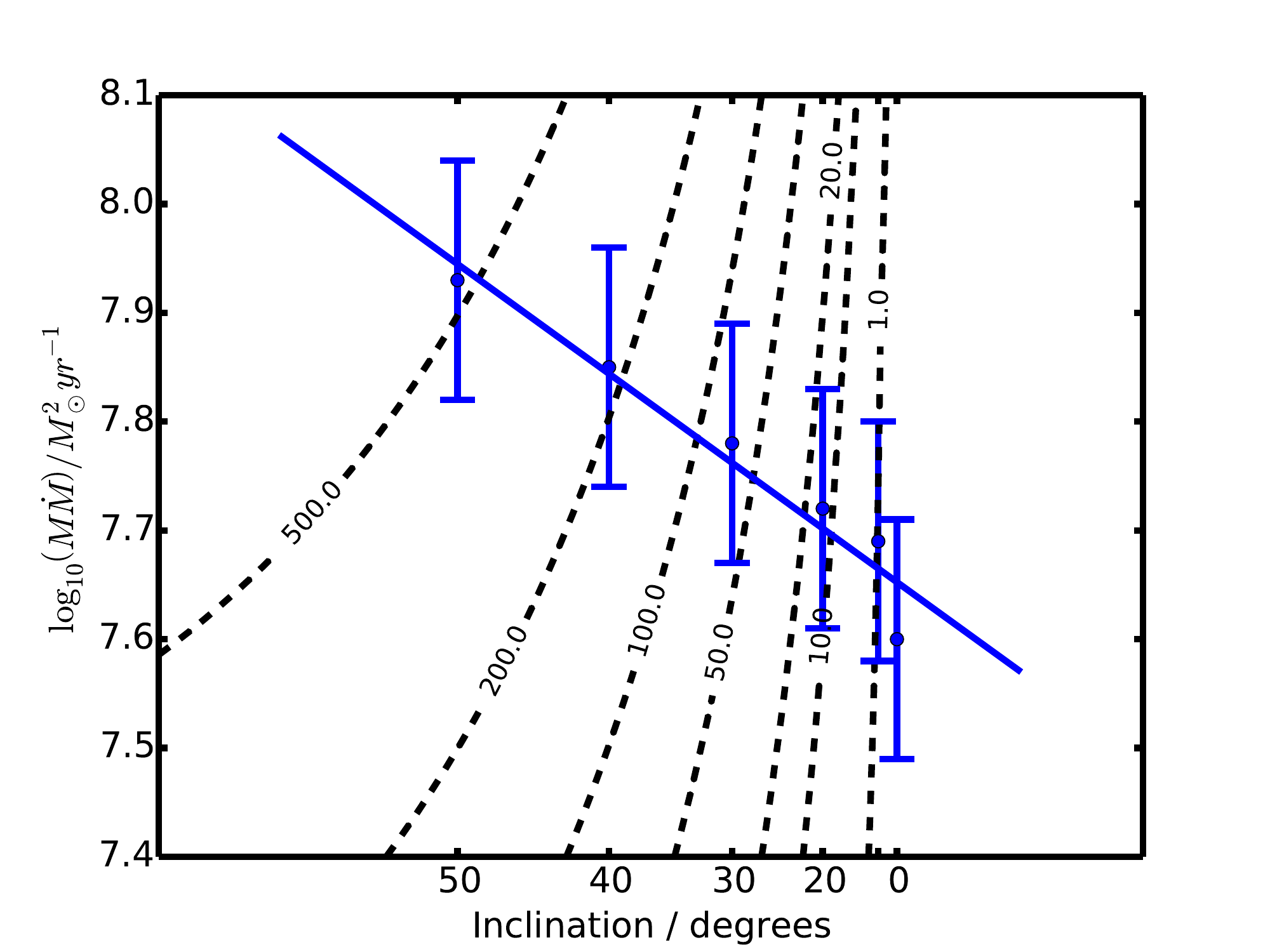}
\caption{$\mmdot$ parameters with uncertainties plotted vs. assumed inclination.  Contours show constant Eddington ratios evaluated assuming a black hole mass from \citet{pa14}. To calculate the Eddington luminosity for our inclinations, we assume a disc-like BLR with a black hole mass that decreases toward edge-on inclinations as $M_\mathrm{BH}$ = $M( \frac{\sin 50}{\sin i})^2$.} 
\label{fig_bias}
\end{figure}

While CREAM can be used to simultaneously fit both $\mmdot$ and $i$, our data are too sparsely sampled with too little overlap between the X-ray and optical light curves to provide a simultaneous fit.  To remedy this, we fix the inclination and allow the $\mmdot$ parameter to vary.  We do this for inclinations 0 - 50 degrees in 10 degree increments. The CREAM fit for the 50 degree case is shown in Fig. \ref{fig_50deg} with a result of log$\mmdot =7.92 \pm{0.11}$, where M is in units of $M_{\odot}$ and $\dot{M}$ is in $M_{\odot}$~yr$^{-1}$.  We note from Fig. \ref{fig_bias} that lower assumed inclinations result in lower estimates of $\mmdot$.

Modelling of the $H\beta$ emission line in NGC 6814 by \citet{pa14} has allowed for a mass and inclination to be determined for this object, which, in turn allows us to determine the mass accretion rate implied by our best fitting model.  For $i = 50^\circ$, and $M=10^{6.42}$ M$_\odot$, we get $\dot{M} = 31.6$ M$_\odot$ yr$^{-1}$, which, assuming an accretion efficiency $\eta = 0.1$ implies an Eddington fraction of $L_{\rm bol}/L_{\rm Edd} = 546$.  Additionally, standard reverberation analysis gives a black hole mass of $M_{\rm BH} =10^{7.04 \pm 0.06}$ $M_{\odot}$ \citep{bentzDB}, using the weighted virial product of all broad lines from \citet{bentz09b} and the $f$-factor from \citet{grier13}.  For the updated \citet{bentz09b} mass, we get $\dot{M} = 7.6$ M$_\odot$ yr$^{-1}$, and an Eddington fraction of $L_{\rm bol}/L_{\rm Edd} = 31.8$.

\subsection{Multi-component spectral decomposition}

\begin{figure}
\centering
\includegraphics[width=8cm]{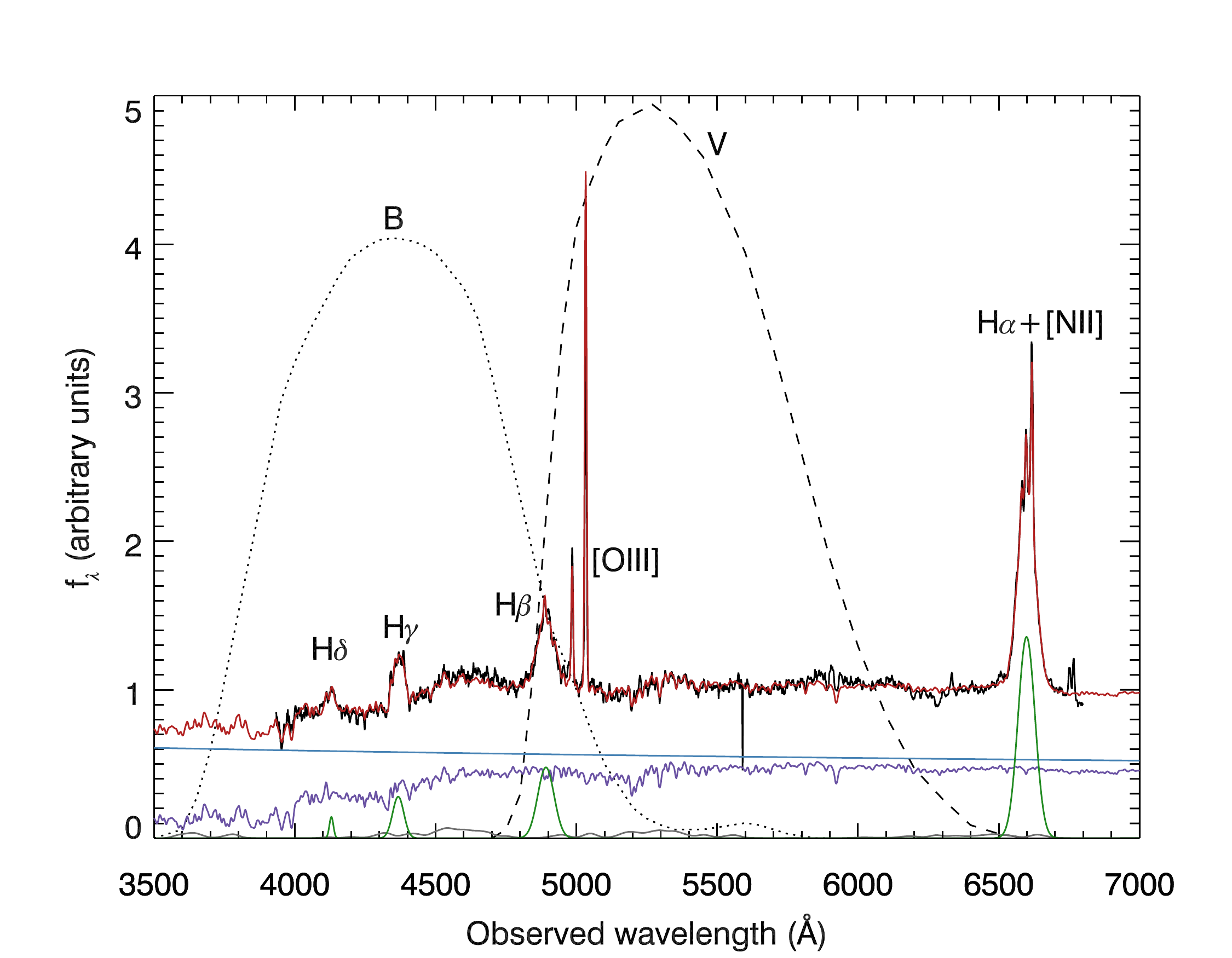}
\caption{The 6dF optical spectrum of NGC~6814 (black).  Dotted and dashed black lines show the transmission curves for the Liverpool Telescope   $B$ and $V$ filters.  The best-fitting composite model is shown in red.  Also shown is the  galactic stellar template (purple), continuum power-law (blue), and broad emission lines (green). }
\label{fig:spectrum}
\end{figure}

We estimate the contribution of the broad lines to the $B-$ and $V-$bands through fitting an archival spectrum of NGC 6814.  We obtained the 6dF spectrum \citep{jones09} of NGC~6814 from NED\footnote{\url{https://ned.ipac.caltech.edu/}}.  We then follow the spectral decomposition method described in \citet{barth13} in order to determine the flux of individual components.  We fit the spectrum with a model consisting of a power-law continuum, galaxy stellar template, \ion{Fe}{ii} template and Gaussians for the broad and narrow emission lines.  The model was convolved with a Gaussian to match the spectral resolution of 6dF.  We use the \ion{Fe}{ii} template of \citet{veroncetty04} convolved with a broad Gaussian, assuming it originates in the BLR. The best-fitting Gaussian width for the \ion{Fe}{ii} complex is consistent with the widths of the broad lines (approximately the same as $H\gamma$, but narrower than $H\alpha$ or $H\beta$).  For the galaxy stellar template we use a model from \citet{maraston11} which assumes a stellar population of 11 Gyr, with solar-abundance, a Saltpeter IMF and uses the MARCS theoretical stellar library.  We fit the model to the data using the non-linear least squares curve fitting package of \citet{markwardt09}.  The spectral fit is shown in Fig. \ref{fig:spectrum}.

Once we obtained a good fit, we calculate the fraction of the flux in both the $B-$ and $V-$bands from each component, weighting the best-fitting spectral model by the transmission curves for each filter.  We compare only the power-law continuum flux with the broad line (including \ion{Fe}{ii}) flux in each filter, since the galaxy and narrow line fluxes will remain constant.  We find that broad lines are approximately 9\% and 8\% of the AGN flux in the $B-$ and $V-$bands, respectively.

\section{Discussion}
We observed the AGN NGC~6814 for approximately 100 days with \textit{Swift} and 70 days with the Liverpool Telescope, obtaining X-ray, UV and optical lightcurves.  The lightcurves are all strongly correlated, with the X-ray lightcurve showing the sharpest variability features and highest variability amplitude. Cross-correlation analysis shows that the UV and optical bands lag behind the X-ray by approximately 2 days.  The lags, variability amplitude and the smoothing of longer wavelength lightcurves are all consistent with a scenario where the X-rays irradiate, and are reprocessed in, the accretion disc to drive the UV/optical variability.

To investigate this scenario further, we fit the lightcurves using CREAM, a MCMC code that assumes a standard thin disc irradiated by the X-ray source.  This model fits the data well, allowing us to constrain the product $M\dot{M}$. Using two different estimates of black hole mass, we calculated mass accretion rates and corresponding Eddingtion fractions, finding highly super-Eddington fractions. Based on the observed flux from NGC 6814, such highly super-Eddington accretion is clearly not occurring.  The average host-galaxy subtracted $V-$band flux density is approximately $5.9\times10^{-16}$ erg cm$^{-2}$ s$^{-1}$ \AA$^{-1}$.  We use this to estimate the bolometric luminosity of NGC 6814 during our observations.  We do this by assuming that the $V-$band flux density is approximately the flux density at 5100\AA.  We then apply an extinction correction assuming $E(B-V) = 0.1586$ (the \citealt{schlafly11} corrected value from \citealt{schlegel98}) and the extinction law of \citet{cardelli89}.  We calculate the luminosity distance assuming a cosmology of $H_0 = 70$ km s$^{-1}$ Mpc$^{-1}$, $\Omega_M = 0.3$, $\Omega_\Lambda = 0.7$.  We then apply a bolometric correction assuming $L_{\rm bol} = 9\lambda L_\lambda(5100\AA)$ (while there are more nuanced bolometric corrections, this is sufficient for our basic estimate here).  Doing this gives an estimated $L_{\rm bol} = 2.7\times10^{42}$ erg s$^{-1}$, which, corresponds to $L_{\rm bol}/L_{\rm Edd} = 0.008$ for the \citet{pa14} mass, and 0.002 using the updated \citet{bentz09b} mass.  Since $\tau \propto \dot{M}^{1/3}$, decreasing the mass accretion rate by a factor of $546/0.008$ or $31.8/0.002$ (depending on the mass assumed), would lead to predicted lags a factor of about 40 or 25 smaller, respectively.  In other words, for realistic values of mass and mass accretion rate, the observed lags are significantly longer than predicted by the standard thin disc model and hence the UV/optical emitting region is further out.

This discrepancy between standard disc model and observed lags is common among AGN where wavelength-dependent lags have been observed.  In \citet{troyer_16}, a standard thin-disc model was fit to the lags and fluxes of a sample of 14 AGN and used to estimate the distances to those objects.  However, the measured distances implied $H_0 = 44\pm5$ km s$^{-1}$ Mpc$^{1}$, a factor of 1.6 smaller than the generally accepted value.  This is a different manifestation of the problem.  The model used by \citet{troyer_16} has $D \propto \tau \lambda^{-3/2}f_\nu^{-1/2}$.  Since $H_0 \propto 1/D$, the discrepancy with $H_0$ implies that the observed lags are too large by a factor of 1.6 on average.  

More recently, wavelength-dependent lags in NGC~5548 measured from long-term monitoring campaigns in 2013 and 2014 also show that while the lags follow the expected $\tau\propto\lambda^{4/3}$ dependence, they are also larger than expected given reasonable values for mass and mass accretion rate \citep{troyer_18,troyer_39,faus_15}.  For instance, \citet{troyer_18} have to increase $M\dot{M}$ by a factor of 3, as well as change other parameters in their model, in order to get good agreement with the lags.  \citet{troyer_39} compare both the wavelength-dependent lags in NGC 5548 and the lags in NGC 2617 measured by \citet{troyer_20} with predictions based on reasonable $M\dot{M}$ for those objects, again showing that both exhibit longer lags than expected.  In MCG-6-30-15  \citet{lira15} also find larger than expected lags, showing that only with an unreasonable increase in X-ray luminosity (a factor of 4 higher) will the measured lags be in good agreement with theory.

\citet{troyer_18} and \citet{troyer_39} note that this discrepancy with the standard thin disc model is consistent with the results from gravitational microlensing, which have also found that the UV and optical emitting regions seem to be further out than predicted by the standard thin disc model \citep[see][and references therein]{mosquera13,blackburne15}.  One possible explanation for this difference is that the accretion disc is inhomogeneous, with many different zones whose temperatures vary independently \citep{troyer_40}.  In this model, the global time-averaged properties of the disc follow the standard thin disc temperature profile, however, instabilities in the disc can lead to local zones whose temperature varies. With a large enough number of zones and amplitude of temperature fluctuations, the half-light radius of the disc increases enough to match the observed microlensing results.  This is just one of several scenarios discussed in the literature, and we refer the reader to other detailed discussions on this discrepancy  \citep[see][and references therein for detailed discussions]{troyer_16,troyer_40,troyer_18,troyer_39,lira15,faus_15}.

The lags in NGC 5548 are the best constrained for any source thus far, so provide an interesting comparison to our results on NGC 6814.  From the standard thin disc model, we would expect lags to scale like $(M\dot{M})^{1/3}$.  The mass and mass accretion rate for NGC 6814 are both estimated to be smaller than for NGC 5548.  Using the \citet{bentz09b} and \citet{pa14} masses and the estimated mass accretion rates given above, we would expect the lags in NGC~6814 to be about a factor of $10-17$ smaller than NGC~5548, yet, the lags are comparable between the two sources.  The reason for the difference is not clear, and our interpretation is limited by the fact that the $B-$ and $V-$ band lags are not well constrained in NGC~6814.  Future monitoring utilizing more wavebands and achieving better constrained lags could help understand the differences.

Since the lags are measured using broadband photometric filters, broad emission lines falling within the filter can increase the measured lag \citep{chelouchezucker13, ch13}.  We can  do a simple estimate of this for NGC 6814 by considering the broad line contamination.  If we assume a 1.5 day continuum lag, and a BLR lag of 7 days \citep[the H$\beta$ lag for NGC~6814 is approximately this value;][]{bentz09b}, with 9\% of the flux originating in the BLR implies an observed lag of: $\tau = 0.91 \times 1.5$ days + $0.09 \times 7$ days = 2.0 days.  Hence, the contribution from broad emission lines may increase the observed lag by 0.5 days, and the lags in $B$ and $V$ could be 25\% smaller than measured (though note that this is smaller than the size of the uncertainties in the lags).  In addition to contamination from broad emission lines, diffuse continuum emission from broad-line clouds can also contaminate the lags.  \citet{troyer_41} show that reflected and thermal diffuse continuum can broadly mimic the $\tau\propto\lambda^{4/3}$ dependence, and may account for about one-third of the lag between 1350\AA\ and 5100\AA.  UV and optical \ion{Fe}{ii} pseudo-continuum emission from BLR clouds or an intermediate region between the accretion disc and the BLR may also contribute \citep{troyer_39}.  Future spectroscopic measurements of wavelength-dependent lags to avoid BLR contamination in AGN will produce more accurately constrained continuum lags and help us further understand the structure of the accretion disc.

\section*{Acknowledgements}
MCB and JES gratefully acknowledge support through NSF CAREER grant AST-1253702 to Georgia State University.  We thank the {\it Swift} team for scheduling and performing the monitoring campaign.  The Liverpool Telescope is operated on the island of La Palma by Liverpool John Moores University in the Spanish Observatorio del Roque de los Muchachos of the Instituto de Astrofisica de Canarias with financial support from the UK Science and Technology Facilities Council.

\bibliographystyle{mnras}
\bibliography{agn}

\bsp	
\label{lastpage}
\end{document}